%
%
%

\def\spose#1{\hbox to 0pt{#1\hss}}
\def\lta{\mathrel{\spose{\lower 3pt\hbox{$\mathchar"218$}}
     \raise 2.0pt\hbox{$\mathchar"13C$}}}
\def\gta{\mathrel{\spose{\lower 3pt\hbox{$\mathchar"218$}}
     \raise 2.0pt\hbox{$\mathchar"13E$}}}

\input psfig.sty

\ifx\mnmacrosloaded\undefined \input mn\fi

%

\newif\ifAMStwofonts

\ifCUPmtplainloaded \else
  \NewTextAlphabet{textbfit} {cmbxti10} {}
  \NewTextAlphabet{textbfss} {cmssbx10} {}
  \NewMathAlphabet{mathbfit} {cmbxti10} {} 
  \NewMathAlphabet{mathbfss} {cmssbx10} {} 
  \ifAMStwofonts
    \NewSymbolFont{upmath} {eurm10}
    \NewSymbolFont{AMSa} {msam10}
    \NewMathSymbol{\upi}     {0}{upmath}{19}
    \NewMathSymbol{\umu}     {0}{upmath}{16}
    \NewMathSymbol{\upartial}{0}{upmath}{40}
    \NewMathSymbol{\leqslant}{3}{AMSa}{36}
    \NewMathSymbol{\geqslant}{3}{AMSa}{3E}

  \else
    \def\umu{\mu}
    \def\upi{\pi}
    \def\upartial{\partial}
  \fi
\fi


\pageoffset{-2.5pc}{0pc}

\loadboldmathnames



\pagerange{00--00}    
\pubyear{1999}
\volume{000}

\begintopmatter  

\title{Collisional baryonic dark matter halos}

\author{Mark A. Walker}

\affiliation{Special Research Centre for Theoretical Astrophysics, School of Physics,
University of Sydney, NSW 2006, Australia}

\shortauthor{Walker}
\shorttitle{baryonic halos}


\acceptedline{Accepted \ \ \ \ \ \ \ \ \ \ \ \ \ \ \ \ \ \ \ \ \ \ . Received}

\abstract{If dark halos are composed of dense gas clouds, as has
recently been inferred, then collisions between clouds lead
to galaxy evolution. Collisions introduce a core in an initially
singular dark matter distribution, and can thus help to reconcile
scale-free initial conditions -- such as are found in simulations
-- with observed halos, which have cores. A pseudo-Tully-Fisher
relation, between halo circular speed and visible mass (not luminosity),
emerges naturally from the model: $M_{vis}\propto V^{7/2}$.

Published data conform astonishingly well to this theoretical prediction.
For our sample of galaxies, the mass-velocity relationship has much less
scatter than the Tully-Fisher relation, and holds as well for dwarf
galaxies (where diffuse gas makes a sizeable contribution to the total visible
mass) as it does for giants. It seems very likely that this visible-mass/velocity
relationship is the underlying physical basis for the Tully-Fisher relation,
and this discovery in turn suggests that the dark matter is both baryonic
and collisional.
}


\maketitle  

\section{Introduction}
A great variety of dark matter candidates exist, motivated by
diverse pieces of evidence, typically indirect (see, for example,
the review by Trimble 1987). One such piece of
evidence comes from the ``Extreme Scattering Events'' (ESEs: Fiedler
et al. 1987); these are radio-wave lensing events caused by dense
blobs of plasma crossing the line-of-sight. Walker \& Wardle (1998a)
presented a model in which these events are caused by ionised material
associated with planetary-mass, molecular gas clouds in the
Galactic halo. This model is good at explaining the ESE
phenomenon, but carries with it the implication that most of the mass
of the Galaxy is in this cold, dense form. If the Galactic dark matter is
really in cold gas clouds -- as, in fact, has been proposed previously
by a number of authors (Pfenniger, Combes \& Martinet 1994; de Paolis
et~al 1995; Gerhard \& Silk 1996) -- then consistency with a variety
of data requires that these clouds satisfy several constraints (Gerhard
\& Silk 1996). Foremost amongst these is the requirement that collisions between
clouds should not entirely deplete the halo of its dark content. In this
paper we use the simplest halo model, an isothermal sphere, to show
how cloud-cloud collisions lead directly to structural changes in the
halo (\S2), and how this process leads to predictable visible galaxy
masses (\S3); however, no attempt is made to predict the form which
the visible mass should assume (stars vs. diffuse gas).

\beginfigure*{1}
\centerline{\psfig{figure=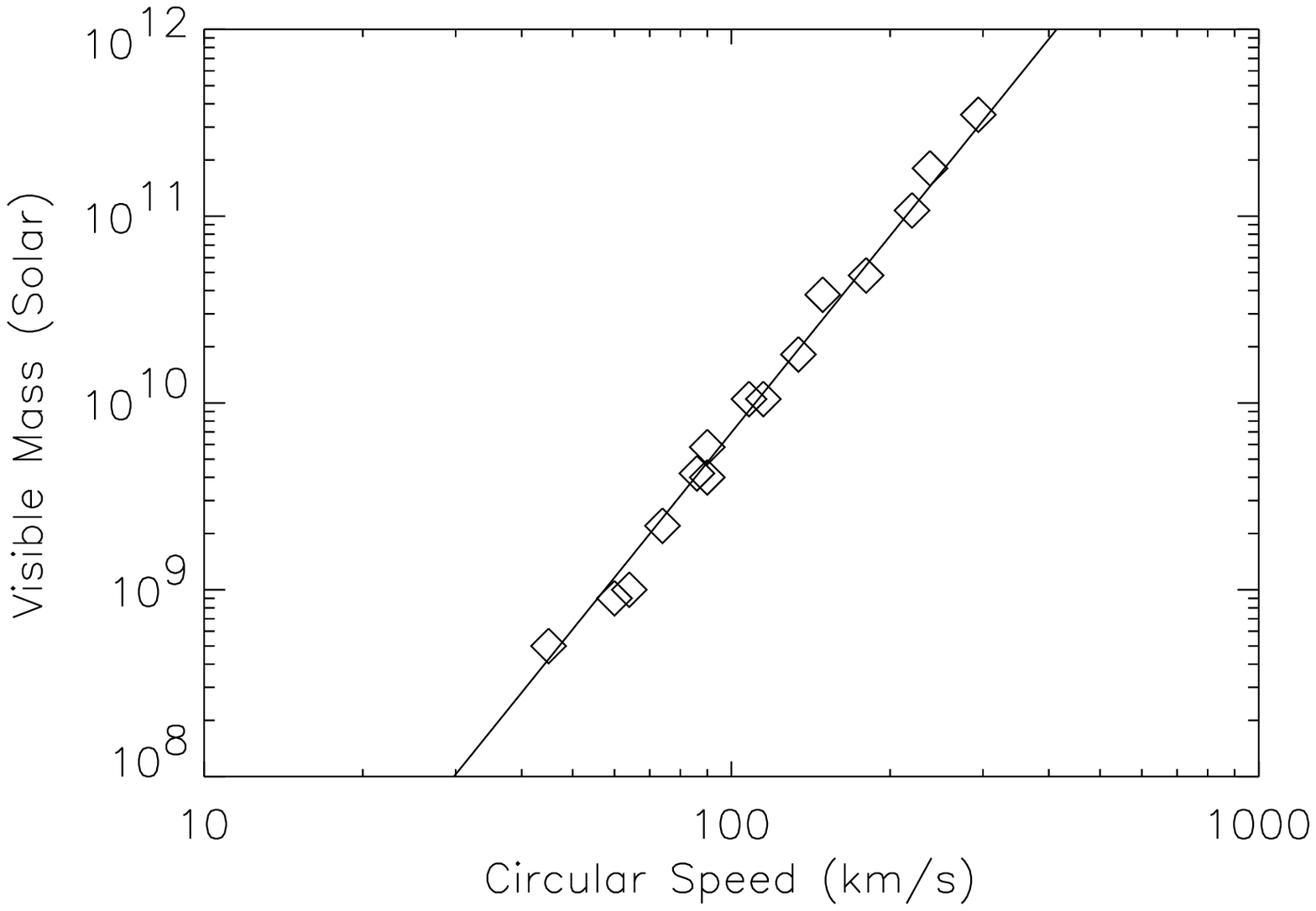,width=14cm}}
\caption{{\bf Figure 1.} Data for a sample of 15 galaxies with well
determined rotation curves (see Table 1), showing the total visible
mass as a function of galaxy circular speed (i.e. $M_{vis}[V]$).
Contributions to $M_{vis}$ come from stars -- both bulge and disk
(determined from ``maximum disk'' decompositions) -- and diffuse gas.
The plotted symbol size is arbitrarily chosen. The line shows the
prediction of equation 9, with cloud surface density ($\Sigma$)
treated as a free parameter.}
\endfigure
\section{Isothermal halos}
In a halo with one-dimensional velocity dispersion $\sigma$ the typical
relative speed of a pair of clouds is $\sqrt{6}\sigma$. Essentially all
collisions are highly supersonic and even glancing impacts -- i.e. those
with impact parameter roughly equal to twice the cloud radius -- are expected
to unbind the clouds, with the result that the collision products
become visible as diffuse gas (which may subsequently be transformed
into stars). In this circumstance the cross-section for disruptive
collisions between clouds is just four times the geometric
cross-section of a single cloud. Recognising that each collision
disrupts two clouds, we see that the collision rate can be written as
$$
-{{\rm d\log_e\rho}\over{{\rm d}t}}=8\sqrt{6}
{{\rho\sigma}\over\Sigma},\hfill\stepeq
$$
where $\rho$ is the halo density, and $\Sigma$ is the mean
surface-density of the individual clouds. A more rigourous treatment,
in which one integrates the collision rate over the velocity distribution
of the clouds, yields a numerical factor ($32/\sqrt{\pi}$) which
differs by less than 10\% from that in eq. 1.

Collisions occur preferentially between pairs of clouds having anti-parallel
velocities, so that the collision products have a much smaller velocity
dispersion than the parent clouds (by a factor of $\sqrt{8}$), and
subsequently undergo infall in the gravitational potential. This infall
modifies the potential, which in turn leads to evolution of the dark halo
density beyond that described by equation 1. (This is ``halo compression''
--- see Blumenthal et al 1986.) We shall not attempt a self-consistent
treatment but, rather, {\it we neglect the evolution of the gravitational
potential.\/} This allows us to estimate the dark halo density evolution
straightforwardly by integrating equation 1; the result is
$$
\rho(r,t) = {{\rho(r,0)}\over{1+t/t_c(r)}}.\hfill\stepeq
$$
where the characteristic time $t_c(r)=\Sigma/8\sqrt{6}\sigma\rho(r,0)$.
If we now model the initial distribution of dark matter as
a singular isothermal sphere, i.e. $\rho(r,0)=\sigma^2/2\pi Gr^2$,
for radius $r$, then equation (2) implies
$$
\rho(r,t) = {{\sigma^2}\over{2\pi G(r^2+r_c^2)}},\hfill\stepeq
$$
where
$$
r_c^2={{4\sqrt{6}}\over{\pi}}{{\sigma^3 t}\over{G\Sigma}}.\hfill\stepeq
$$
That is, at $t>0$ the halo is a non-singular isothermal sphere with core
radius $r_c$. We can recast equation 4 in terms of the native circular speed,
$V$, for the halo, using $V=\sqrt{2}\sigma$. Taking the proposed mean column
for individual ESE clouds (Walker \& Wardle 1998a; see also \S4) of
$N\sim10^{25}\;{\rm cm^{-2}}$, i.e. $\Sigma\sim30\;{\rm g\;cm^{-2}}$,
together with $t=10$~Gyr, we obtain
$$
r_c\sim4.2\,V_{100}^{1.5}\quad{\rm kpc},\hfill\stepeq
$$
where $V=100\,V_{100}\,{\rm km\,s^{-1}}$. In consequence, at the present
epoch we should expect galaxy halos to possess modest cores, if those halos
virialised at redshifts $z\gta1$ (i.e. $\sim10$~Gyr ago). These core radii ought,
in principle, to be measurable from rotation curves of spirals.

In practice the task of measuring a core in the density profile of a 
dark halo is made difficult by the visible galaxy, which itself
contributes to the rotation curve, and one is obliged to perform a
disk-halo decomposition --- a process which engenders a number
of uncertainties. Furthermore, the visible galaxy can alter the
gravitational potential sufficiently that the halo density distribution
changes in response: ``halo compression''; see Blumenthal et al.
(1986). Notice that this effect is {\it not\/} accounted for in
equation 5. These difficulties are mitigated to a large extent
if the visible galaxy is so puny that it acts only as a kinematic
tracer, allowing the rotation curve to be measured but not playing any
significant r\^ole in determining its form. Such galaxies exist and have
already been used to demonstrate that, at least in these cases, halos
{\it do\/} possess cores (Moore 1994; Flores \& Primack 1994). This is
an important result because it appears to conflict with simulations of
halo formation using dissipationless dark matter: these exhibit singular
density profiles (Dubinski \& Carlberg 1991; Navarro, Frenk \& White 1996;
for an alternative view see Kravtsov et al. 1998).
\beginfigure*{2}
\centerline{\psfig{figure=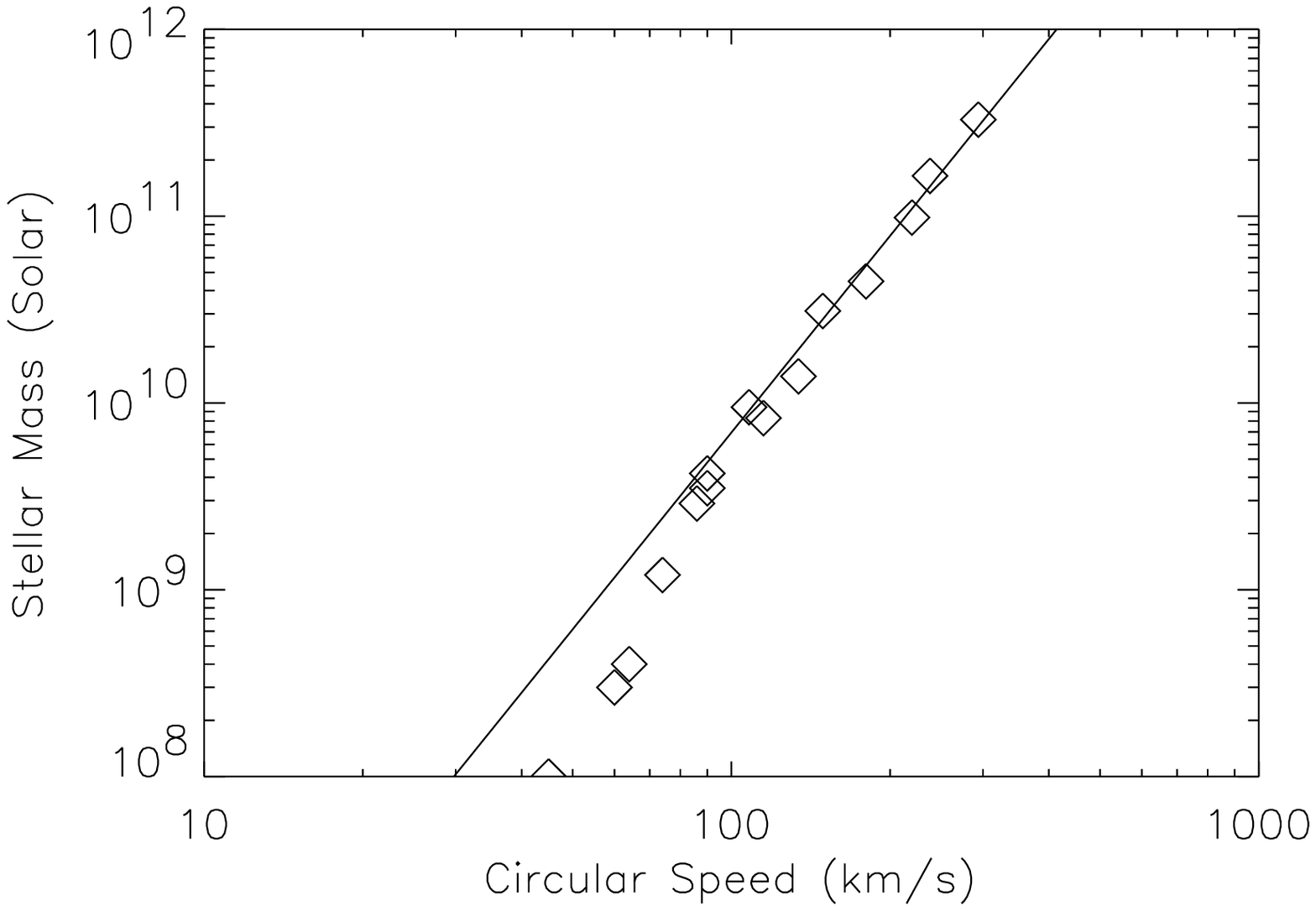,width=14cm}}
\caption{{\bf Figure 2.} As figure 1 but for the stellar mass only,
i.e. excluding the contribution of diffuse gas to the total visible mass;
the theoretical line is identical to that in figure 1. It is clear that
the dwarf galaxies, which have a high mass-fraction in diffuse gas,
systematically depart from the relationship defined by the giants,
if only stellar mass is included in the accounting. Compare
this to figure 1, where a single relation is valid for the whole sample.}
\endfigure
\begintable*{1}
\caption{{\bf Table 1.} Galaxies with well determined rotation curves.
Circular speeds are taken from table A.1 (column 9), p94 of Broeils (1992);
while masses are obtained from ``maximum disk'' rotation-curve decompositions
--- the sum of columns 7 (diffuse gas), 9 (disk stars) and 11 (bulge stars)
of table 2, p244, in Broeils (1992). Only 15 galaxies are common to both tables.}

\halign{
\rm#\hfil&\qquad\rm#\hfil&\qquad\rm#\hfil&#\qquad&\qquad\rm#\hfil
&\qquad\rm#\hfil&\qquad\rm#\hfil&#\qquad&\qquad\rm#\hfil
&\qquad\rm#\hfil&\qquad\rm#\hfil\cr

Name&$V$&$M_{vis}$&&Name&$V$&$M_{vis}$&&Name&$V$&$M_{vis}$\cr
&(${\rm km\;s^{-1}}$)&$(10^{10}\;{\rm M_\odot})$&&
&(${\rm km\;s^{-1}}$)&$(10^{10}\;{\rm M_\odot})$&&
&(${\rm km\;s^{-1}}$)&$(10^{10}\;{\rm M_\odot})$\cr

NGC55&86&0.42&&NGC247&108&1.05&&NGC300&90&0.58\cr
UGC2259&90&0.40&&NGC1560&74&0.22&&NGC2403&134&1.82\cr
NGC2841&294&35.0&&NGC2903&180&4.82&&NGC3109&64&0.10\cr
NGC3198&149&3.80&&DDO154&45&0.05&&NGC5033&220&10.8\cr
DDO170&60&0.09&&NGC6503&115&1.05&&NGC7331&238&18.0\cr
}
\endtable

Kormendy (1992) has derived scaling laws for
real (late-type spiral) galaxy halos; his relationship between core radius
and halo circular speed is (for $H_0=75\;{\rm km\,s^{-1}\,Mpc^{-1}}$)
$$
r_c=3.6\,V_{100}^{1.6}\quad {\rm kpc};\hfill\stepeq
$$
encouragingly close to equation 5. This scaling is less secure than the
simple fact that cores exist in galaxy halos, and one should beware
of undue emphasis on the numerical agreement between equations 5 and 6,
as the mean cloud column density ($\Sigma$) is really only known to
within a factor of a few (see also the comments in \S4). Indeed in the
following section we shall treat $\Sigma$ as a free parameter, and the
resulting best fit indicates a larger value of $\Sigma$ than used here.

\section{A physical basis for the Tully-Fisher relation}
The Tully-Fisher (TF) relation is an empirical result which connects
the width of 21~cm line emission, $\Delta{\cal V}$, from a spiral
galaxy with the galaxy's expected luminosity, $L$:
$$
L\propto\Delta{\cal V}^\alpha,\hfill\stepeq
$$
where $\alpha\simeq4$ (Strauss \& Willick 1995); the scatter in
this result is quite small. Unfortunately there is no sound theoretical
basis for the TF relation. Attempts to derive it on the basis of
the virial theorem, plus the assumption of fixed surface-brightness
for all galaxies (``Freeman's law''), fall foul of the fact that TF
holds also for very low surface-brightness galaxies (Zwaan et~al 1995;
McGaugh \& de Blok 1998). Put another way, we expect that $\Delta{\cal V}
\sim2V$, so that the kinematics reflect the  properties of the dark halo,
but $L$ is a manifestation of the stellar component of the galaxy, and
no direct coupling between these two is expected a priori. Indeed, since
total mass-to-light ratios can vary by more than
an order of magnitude amongst the population of observed galaxies, one
tends to think of the visible and dark components as almost independent.
However, the collisional process we have described in \S2 converts dark
matter into visible forms, and thus creates a close coupling between
the dark halo and the visible galaxy. Such a coupling holds promise for
explaining the Tully-Fisher relation; this appealing attribute of baryonic
models has been recognised previously (Pfenniger, Combes \& Martinet 1994;
Gerhard \& Silk 1996).

We can compute the total visible mass, $M_{vis}$, in the form of stars and gas
(without actually saying anything about their relative proportions), from
$$
M_{vis}(t)=\int_0^\infty{\rm d}r\,4\pi r^2[\rho(r,0)-\rho(r,t)].\hfill\stepeq
$$
With the evolving isothermal halo described by equation 3, this gives
$$
M_{vis}={{\pi\sigma^2}\over G} r_c.\hfill\stepeq
$$
Notice that $M_{vis}$ is perfectly well defined, despite the total mass of the
dark halo diverging at large radii. If we adopt the reasonable supposition of a
roughly similar {\it stellar\/} mass-to-light ratio for all spirals then, provided
that diffuse gas makes a negligible contribution to $M_{vis}$, we expect
$L\propto V^{3.5}$, which is close to equation 7. In the usual form, however,
TF is a relation between global 21~cm line-width, $\Delta{\cal V}$, and luminosity.
Moreover, various exponents $\alpha\simeq4$ are observed, so it is not
immediately obvious that our theory is in agreement with the data.

Spectral line imaging (notably 21~cm imaging) gives detailed information
on the velocity field of a galaxy, allowing accurate determination of the
rotation curve (hence $V$), rather than just the global property $\Delta{\cal V}$
which can be determined with a single dish radio telescope. Broeils (1992,
chapter 4, appendix A) has investigated the $L[\Delta{\cal V}]$ and $L[V]$
relationships for a sample of 21 galaxies with well determined rotation curves.
He finds that blue luminosity is more tightly correlated with $V$ than with
$\Delta{\cal V}$; the former displays a scatter of 0.22, and the latter 0.28,
in $\log_{10}L$. (More precisely, Broeils used the circular speed in the flat
part of the rotation curve, which we take to be a good estimator for $V$.)
Furthermore, the relation he derives is $L\propto V^{3.4}$, which is very
similar to the form we predict.

We could go further and compare the normalisation
of this result with that of our own theory, but in doing so we would be
forced to introduce an additional quantity, the value of the stellar
mass-to-light ratio, which is a priori unknown. This can be avoided
if we compare $M_{vis}$ directly with mass estimates derived from rotation
curve decompositions which, in effect, measure the stellar mass-to-light
ratio for each galaxy. Furthermore, by utilising mass, rather than
luminosity, we expect better agreement with our theory because we can
include the diffuse gas content. Note that diffuse gas is often a
substantial fraction of the total visible mass of dwarf galaxies; in
the (extreme) case of DDO154 it amounts to 80\% of the visible mass
(Carignan \& Freeman 1988). While Broeils (1992) did not investigate
the possibility of any correlation between $M_{vis}$ and $V$, he did
give maximum disk decompositions of the rotation-curves for 15
of the 21 galaxies studied; his results are summarised in table 1,
and plotted in figure 1. Also shown in figure 1 is the relation
given by our equation 9, with the cloud surface density ($\Sigma$)
treated as a free parameter; the fit is manifestly good.

The very first point which must be made about this result is that it
is {\it not\/} simply an artefact of (maximum-disk) rotation curve
decomposition. This point is clear because rotation speed measures
$M/R$, not $M$, so even if the visible mass were the dominant component
we would still be obliged to explain the existence of a characteristic
length scale having a form similar to that of equation 4. However, in the flat
part of the rotation curve -- which has been used to estimate $V$ -- the
visible matter contributes less than the dark matter in every case
in table 1. On average the dark matter contributes three quarters
of the total, for this sample, and for some galaxies the visible
contribution is negligible. The hypothesis that figure 1 simply
reflects the rotation curve decomposition procedure is therefore
untenable. 

The explicit numerical form of our fit is
$$
M_{vis}=7.0\times10^9\,V_{100}^{7/2}\qquad{\rm M_\odot},\hfill\stepeq
$$
corresponding to $\Sigma\simeq140\;{\rm g\,cm^{-2}}$ (for $t=10$~Gyr), and
$$
r_c=1.9\,V_{100}^{3/2}\qquad{\rm kpc}.\hfill\stepeq
$$
We will not attempt to give a figure of merit for the fit quality, as the
uncertainties associated with the data involve systematic uncertainties in
rotation-curve decomposition, and these are hard to quantify. We can, however,
measure the scatter of the data about the theoretical prediction: the
root-mean-square deviation is 0.084 in $\log_{10}M_{vis}$. By comparison the
scatter in the Tully-Fisher relation for this sample is almost a factor of 3
larger (0.24 in $\log_{10}L_B$ --- slightly bigger than the scatter of 0.22
for the full sample of 21 galaxies studied by Broeils). Notice that only a
part of this difference can be accounted for by uncertainties in galaxy
distances, as the stellar mass inferred from rotation curve decomposition
is proportional to distance, while luminosity and the measured gas mass
are both proportional to (distance)$^2$. We also note that if the mass
contributed by diffuse gas is neglected then the dwarf galaxies systematically
depart from the relation defined by the giants; this point is graphically
illustrated in figure 2, where stellar mass is plotted as a function of halo
circular speed. {\it These facts oblige us to conclude that the fundamental
connection is between total visible mass and halo circular speed, with
the Tully-Fisher relation emerging as an approximation which is valid
when most of the visible mass is in stellar form.\/}

To add a little more weight to this conclusion, we emphasise that
our sample of galaxies is very heterogeneous: it spans the entire size
spectrum from dwarfs to giants; it includes low surface-brightness
objects; most importantly, perhaps, it includes cases in which the
visible galaxy makes a negligible contribution to the rotation curve.
This last point is crucial as it requires that the observed $M_{vis}[V]$
relation be interpreted as $M_{vis}$ being determined by $V$, and not
the other way around. That is, purely from an observational perspective
we can assert that the visible mass content of a galaxy is
determined by the velocity dispersion of the dark matter halo.
Our theory shows why this ought to be so, and it follows that these
data support the model of a baryonic dark halo, with clouds of
typical surface density $\Sigma\simeq140\;{\rm g\,cm^{-2}}$.

\section{Discussion}
The evolution implicit in equation 9
could, in principle, be used to further test the theory we have presented,
but in practice this will be difficult. In particular the requisite
sensitivity and angular resolution, for determining 21~cm rotation curves
of normal galaxies at $z\sim1$, are both beyond the reach of current
instrumentation. Some studies have already been made of Tully-Fisher-type
correlations of galaxies at $z\sim1$, based on optical data alone
(e.g. Vogt et al. 1997). These show mild evolution in a sense opposite
to that expected in our theory (if a constant mass-to-light ratio is
assumed), and we must suppose that this is a result of stellar populations
being younger at earlier epochs. Notice that there are serious consequences
for galaxy distance estimates based on a {\it local\/} Tully-Fisher
relation if the actual TF relation evolves with look-back time.

It is well known that the infrared TF relations show less scatter than
their visible counterparts (Aaronson, Huchra \& Mould 1979), because infrared
photometry is less sensitive to the young, hot stars which tend to
dominate the luminosity (but not the total stellar mass) of the galaxy.
Extinction corrections are also smaller in the infrared. This argues
that rotation-curve decompositions based on infrared photometry ought
to be better than those based on visible photometry. In addition,
there is no particular reason to suppose that ``maximum disk'' rotation curve
decompositions give the best estimates of the stellar mass-to-light ratio. Thus
with a study that is specifically aimed at testing equation 9, it may be
possible to reduce the scatter seen in figure 1 below its already small value.
Ideally one would like to work with a sample of galaxies from a single
cluster, thereby reducing the dispersion contributed by distance uncertainties;
distance errors probably dominate the currently observed scatter of 0.084 dex in
the $M_{vis}[V]$ relation.

Any cluster of galaxies which is well approximated by an
isothermal distribution of dark matter should fit into the scheme we
have outlined, simply by an appropriate choice of velocity dispersion.
(We estimate from the evaporation constraints given by Gerhard \& Silk
[1996] that the clouds can survive for roughly a Hubble time in the
environment of a rich cluster of galaxies.) For example, taking
$\sigma=10^3\;{\rm km\,s^{-1}}$ equation 10 predicts a visible mass of
$M_{vis}=7\times10^{13}\;{\rm M_\odot}$, broadly consistent with the
data (Jones \& Forman 1984), and equation 11 gives a dark matter core radius
of $r_c=100\;{\rm kpc}$. This core radius is rather larger than indicated
by analysis of cluster lensing data (Miralda-Escud\'e 1993; Flores \& Primack
1994) but we must bear in mind that lensing measures the total surface density,
not just the dark component. For galaxies the greatest visible contribution
typically comes from stars, whose presence is relatively straightforward
to quantify, but for rich clusters it is usually the hot intra-cluster
medium which dominates, and here the inferred mass distribution is much
more model dependent. (Notice, again, that the theory we have presented is
only partially predictive in that it gives the total visible mass, but does not
tell us whether this ought to be in stars or diffuse gas.) In particular we note
(i) it is thought that in ``cooling flow'' clusters a large amount of gas accumulates
in some (unknown) form at the centre of the cluster (Fabian 1994), and
(ii) there are tentative detections of huge amounts of warm (EUV emitting)
gas in some clusters (Mittaz, Lieu \& Lockman 1998).

In the foregoing discussion we have given no consideration to the
material properties of the cloud collision products, being content
with the notion that this stuff becomes part of the visible pool. A basic
analysis of the effects of the shock (Walker \& Wardle 1998b) indicates
that, for the Galactic halo, the result of a collision will typically
be atomic gas, even though the clouds are initially molecular. This suggests
a possible connection with the ``High Velocity Clouds'' (Wakker \& van~Woerden
1997) which are seen (almost exclusively) in 21~cm emission: some HVCs may simply
be material from recent dark cloud collisions. This gas is expected to have such
a low column density that it will be stopped by the diffuse interstellar medium,
at the first encounter with the Galactic plane, thereby contributing to the
assembly of the gaseous disk and, subsequently, the Milky Way.

As well as being relevant to the Tully-Fisher relation, the model we have presented
is germane to the ``Disk-Halo Conspiracy'' (Sancisi \& van~Albada 1987). The
conspiracy is so-called because it seems puzzling that (giant) spiral galaxy rotation
curves should be as flat as they are observed to be, given that the acceleration
is predominantly due to stars at small galactocentric radii, and due to dark matter
at large radii. In the model we have presented, however, the conspiracy is no
surprise; rather it is an innate feature, as dark matter is
converted to visible forms by cloud-cloud collisions. With the model of \S2
we cannot sensibly compute rotation curves -- because we have neglected evolution
of the gravitational potential -- so these statements are
necessarily qualitative. A useful quantitative treatment would
require a self-consistent description of the evolution; more careful consideration
of the initial conditions (dark matter density profile) would also be appropriate.

The principal difficulty with our model is that the fit shown in figure 1
requires $\Sigma\simeq140\;{\rm g\,cm^{-2}}$ (for $t=10$~Gyr), whereas
Kormendy (1992) measures halo core radii which suggest a smaller cloud
surface density, $\Sigma\sim40\;{\rm g\,cm^{-2}}$. Both of these values
are higher than the original estimate (Walker \& Wardle 1998a) based
on scintillation data, and incompatible with each other. Kormendy (1992)
cautions that his derived halo scaling relationships are quite uncertain,
and it is evident from his figure 3 that the scatter in the measurements is
much larger than in our figure 1, so the value $\Sigma=140\;{\rm g\,cm^{-2}}$
is to be preferred. One possible resolution is that our theory is simply too
crude. For example, in the case of spiral galaxies the fact that the visible
matter has angular momentum requires that the dark matter should also have some,
yet we are modelling halos as non-rotating. Perhaps a more refined theory would
yield consistent estimates of halo core radius and visible mass? We have made
a preliminary investigation of this possibility, using Toomre's (1982) analytic
models of rotating isothermal halos. These models span the entire range from
isothermal spheres to cold, rotationally supported disks. We find that that
the prediction $M_{vis}\propto V^{7/2}$ holds for this whole set of models,
with a coefficient of proportionality which is only weakly dependent on the
degree of rotational support that the halo possesses. Similarly the inferred
core radius is only weakly dependent on rotation, so it seems unlikely that
the discrepancy can be explained in this way. The fact that the predicted $M_{vis}$
is only a weak function of the degree of halo rotation, for a fixed value
of $V$ does, however, lend a certain robustness to the model we have presented.

Finally we note that the preferred cloud column density deduced in this paper,
i.e. $\Sigma\simeq140\;{\rm g\,cm^{-2}}$, is substantially larger than the value
$\Sigma\sim30\;{\rm g\,cm^{-2}}$ anticipated on the basis of the ESE data
(Walker \& Wardle 1998a). The latter value was derived from the sky covering
fraction, $\tau$, of the ESE clouds, together with the requirement that the
cold clouds not contribute more than the dynamically determined mass for the
Galactic dark halo. Using the density distribution of equation 3, we find that
at high Galactic latitude,
$$
\tau\simeq{{2.4\times10^{-4}}\over\sqrt{\Sigma_{100}}},\hfill\stepeq
$$
at the Solar circle, where $\Sigma=100\,\Sigma_{100}\;{\rm g\,cm^{-2}}$. Evidently
if $\Sigma_{100}=1.4$, then the dynamical constraints limit the covering
fraction of the clouds to be $\tau\la2\times10^{-4}$.  The observed ESE
cloud covering fraction is highly uncertain because what actually
constitutes an ESE is only very loosely defined. Selecting all periods of
``unusual variability'', Fiedler et al (1994) reported $\tau_{obs}\sim5\times
10^{-3}$. On the other hand, if we consider events whose light-curves
are reminiscent of the calculations of Walker \& Wardle (1998a), then
we would include only the events for the sources 0954+658 and 1749+096,
leading to an estimate $\tau_{obs}\sim5\times10^{-4}$; this is still 2.5
times larger than we expect for $\Sigma=140\;{\rm g\,cm^{-2}}$.

To understand how this discrepancy might arise we need to recognise
that the size of an ESE cloud relates to the inner boundary of the ionised
wind, as it is the free electrons which comprise the radio-wave ``lens''.
By contrast the measure of $\Sigma$ we have given in this paper is a characteristic
surface density for the underlying hydrostatic cloud; the wind is irrelevant here
as it contains a negligible fraction of the mass of the cloud. Walker \& Wardle (1998a)
explicitly assumed that the inner boundary of the ionised wind coincided
with the surface of the hydrostatic cloud. However, if the inner boundary
of the ionised wind is actually the outer boundary of a neutral wind,
then the ESE covering fraction will be substantially larger than the estimate
given by equation 12, as is observed to be the case. It is plausible that a neutral
wind should underly the ionised wind, but this possibility requires careful
consideration in its own right and will not be pursued here. Further constraints
on the column density of the cold clouds are considered by Draine (1998),
based on optical refraction by the neutral gas (``gas lensing''). These
constraints are compatible with $\Sigma\simeq140\;{\rm g\,cm^{-2}}$
provided that the internal density profile of the clouds is not strongly
centrally concentrated; convective polytropes, for example, are acceptable.

\section{Conclusions}
Modelling galaxy halos as isothermal spheres composed of collisional, baryonic
dark matter leads us to expect non-singular dark halo density distributions with
a predictable visible mass content. Both of these expectations are borne out in
the data, with the latter result very likely being the fundamental basis for the
Tully-Fisher relation. Although simple, the theory is remarkably good at predicting
(spiral) galaxy masses across the whole size spectrum from dwarfs to giants. The
evident success of this description of galaxy evolution gives strong support to the
notion that galaxy halos are composed of a vast number of cold, dense,
planetary-mass gas clouds.

\section*{Acknowledgments}
Thanks to Mark Wardle, Jeremy Mould, Bohdan Paczy\'nski, Ken Freeman and
James Binney for their helpful comments. The Special Research Centre for
Theoretical Astrophysics is funded by the Australian Research Council under
its Special Research Centres Program.

\bye